\documentclass{article}

\usepackage{customize}

\begin{document}

\title{A transport method for restoring incomplete ocean current measurements}




\author[1,2]{Siavash Ameli\protect\thanks{Email address: \href{mailto:sameli@berkeley.edu}{\protect\nolinkurl{sameli@berkeley.edu}}}}
\author[1]{Shawn C. Shadden\protect\thanks{Email address: \href{mailto:shadden@berkeley.edu}{\protect\nolinkurl{shadden@berkeley.edu}}}}

\affil[1]{\small\textit{Mechanical Engineering}, \small\textit{University of California}, \small\textit{Berkeley, CA, USA 94720}}
\affil[2]{\small\textit{Department of Mathematics}, \small\textit{University of California}, \small\textit{Berkeley, CA, USA 94720}}

\date{}
\maketitle





\begin{abstract}
Remote sensing of oceanographic data often yields incomplete coverage of the measurement domain. This can limit interpretability of the data and identification of coherent features informative of ocean dynamics. 
Several methods exist to fill gaps of missing oceanographic data, and are often based on 
projecting the measurements onto basis functions or a statistical model. Herein, we use an information transport approach inspired from an image processing algorithm. 
This approach aims to restore gaps in data by advecting and diffusing information of {\em features} as opposed to the field itself. Since this method does not involve fitting or projection, the portions of the domain containing measurements can remain unaltered, and the method offers control over the extent of local information transfer. This method is applied to measurements of ocean surface currents by high frequency radars. This is a relevant application because data coverage can be sporadic and filling data gaps can be essential to data usability. Application to two regions with differing spatial scale is considered. The accuracy and robustness of the method is tested by systematically blinding measurements and comparing the restored data at these locations to the actual measurements. These results demonstrate that even for locally large percentages of missing data points, the restored velocities have errors within the native error of the original data (e.g., $<10$\% for velocity magnitude and $<3$\% for velocity direction). 
Results were relatively insensitive to model parameters, facilitating {\em a priori} selection of default parameters for {\em de novo} applications. 

\end{abstract}



\section{Introduction} \label{sec:introduction}

\begin{figure}[t]
    \centering
    \footnotesize{
    \includegraphics[width=\textwidth]{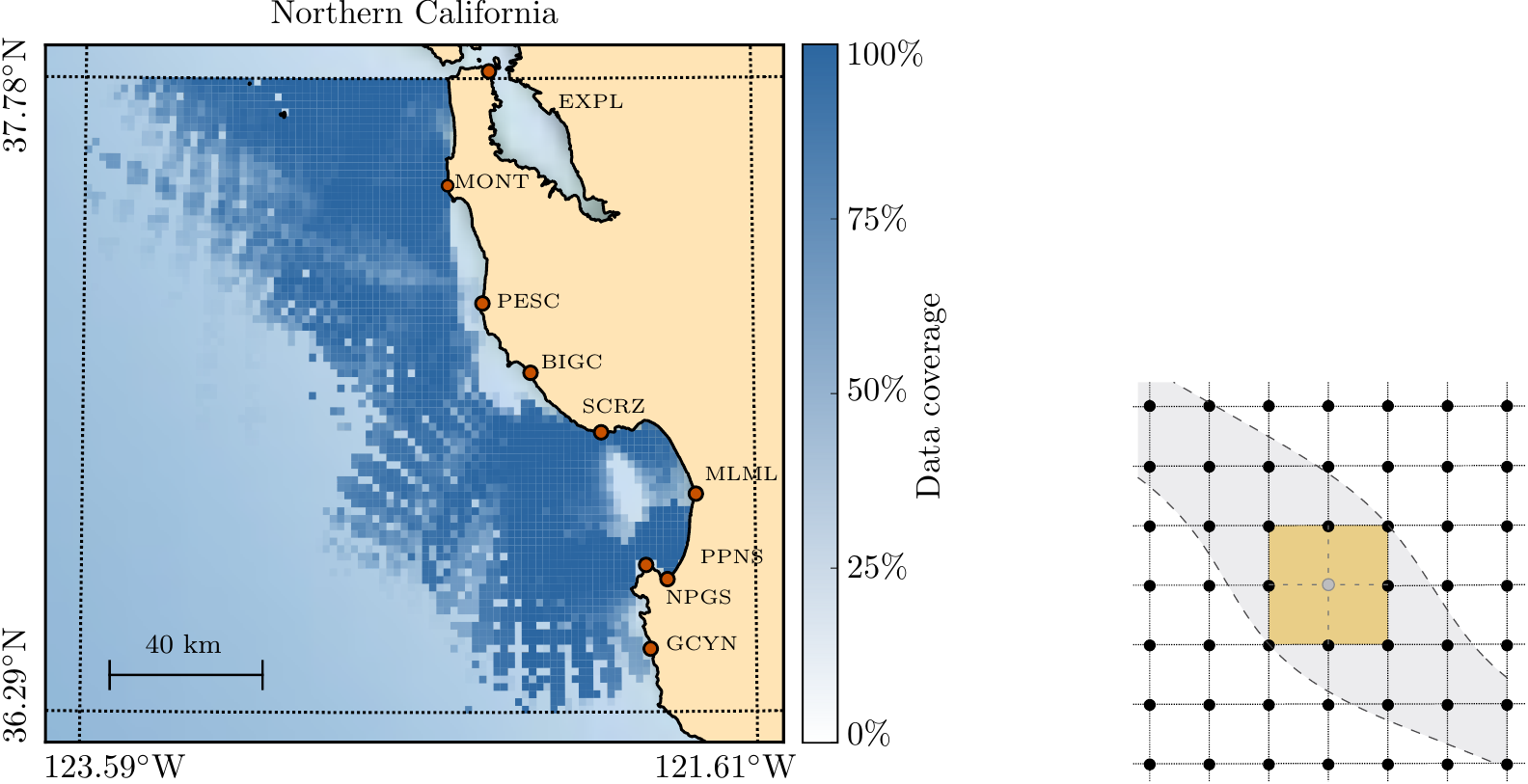}
    }\hspace{0.1\textwidth}
    \def\svgwidth{0.25\textwidth}
    \caption{Left: Percent coverage of radar measurements along the northern California coast, averaged over the month of January 2017. The location of radar sites are indicated along with their code names. Right: While a single missing data point may only affect interpolation in an isolated region (e.g. yellow cells), it can have far reaching effect on tracking trajectories (grey region) potentially passing through this region.} \label{fig:Coverage}
\end{figure}

Measuring ocean surface currents is a common example of remote sensing in oceanography that is prone to incomplete coverage. Remote sensing of coastal surface velocity fields is widely conducted with high frequency (HF) radar techniques~\citep{BARRICK-1977, PADUAN-1997, PADUAN-1999, PADUAN-2009}, whereby a network of land-based radar sites are deployed and each measures the radial component of the surface current from backscatter of the emitted signal. A vector field of the surface velocity can then be reconstructed by combining the overlapping radial measurements from multiple sites using variety of methods (see \eg \citep{LIPA-1983}). Various sources can contribute to an incomplete coverage of measurements, including environmental effects as well as the geometric dilution of precision (GDOP) inherent to the spatial configuration of radar sites~\citep{CHAPMAN-1997, CHAPMAN-1997-2, GRABER-1997-2}. For example, Fig.~\ref{fig:Coverage} (left) shows the average HF radar coverage along the northern California coast from the Coastal Observing Research and Development Center.

A few or isolated missing points might be acceptable for visualization of a field, or an Eulerian analysis that is spatially and temporally localized. However, for applications that leverage Lagrangian analysis, such as the tracking of trajectories or concentration, a complete coverage of vector field is required. Examples of Lagrangian analysis include pollutant tracking \citep{LEKIEN-2005, COULLIETTE-2007}, drifter tracking \citep{OLASCOAGA-2006} and drifter backtracking \citep{BREIVIK-2012}, oil spill evolution \citep{ABASCAL-2009}, coherent structures analysis \citep{SHADDEN-2009, PEACOCK-2013}, search and rescue \citep{ULLMAN-2006,BARRICK-2012}, hazards management \citep{HERON-2016} and data assimilation \citep{SPERREVIK-2017} among others. Since these applications widely rely on the computation of trajectories (of either discrete particles or continuous concentrations), even a few missing points can have a broad effect. Figure \ref{fig:Coverage} (right) illustrates how a single missing data point may affect the tracking of a collection of trajectories passing through any of the four cells containing the missing data point. This issue is amplified if data coverage is scattered, which yields almost unusable data for Lagrangian analysis. Although simple local interpolation can effectively recover isolated missing data points, interpolation can significantly degrade restoration accuracy for larger groups of missing points. Additionally, interpolating missing points in proximity of the ``open boundaries'' (non-coastline boundaries) is often infeasible due to the typical dispersion of missing data (see Fig.~\ref{fig:Coverage}). 

Various methods have been applied to \emph{reconstruct} the velocity vector field from radially based remote measurements as well as to \emph{restore} missing data, which is to overcome coverage limitations. Some methods address these two steps together. A variety of optimal fitting methods have been proposed. A common method is unweighted least square fitting \citep{LIPA-1983}, which combines the fields of radial surface currents to produce a vector field on a fixed grid by assuming that the data are radially uniform. An optimal interpolation technique for oceanographic data was developed \citep{BRETHERTON-1976, DENMAN-1985} by using the Gauss-Markov theorem that yields a least square error of the linear estimate of measurement variables. The penalized least square method based on the 3D discrete cosine transform has been used \citep{FREDJ-2016}. Objective mapping of data with least square fitting has been studied \citep{DAVIS-1985, KIM-2007}. A stable optimal interpolation technique was introduced in \citep{KIM-2008} to account for spatial correlation of data and covariance of uncertainty. Other approaches have applied empirical orthogonal function (EOF) analysis \citep{BOYD-1994, BECKERS-2003, STOKES-2004, ALVERA-2005}, which projects the observations onto the dominant modes of the sample covariance matrix. In \citep{FROLOV-2012} a linear autoregression model is used to predict the temporal dynamics of EOF coefficient, which improves the surface current predictions of HF radar observations. The EOF method was combined with variational interpolation in \citep{YAREMCHUK-2009, YAREMCHUK-2011} to penalizes the spatial variability of error variance of the divergence and vorticity fields. The regularized expectation minimization method was used in \citep{SCHNEIDER-2001}, which iteratively estimates mean values and covariance.
Another class of interpolation methods is based on modal analysis of the domain geometry \citep{LIPPHARDT-2000}, which was extended to the open-boundary modal analysis (OMA) method \citep{LEKIEN-2004, KAPLAN-2007, BARRICK-2012, LEKIEN-2009}. 
Recently an artificial neural network has been used to interpolate data gaps \citep{DAUJI-2016}.

A common objective of interpolation methods is to minimize the error of the interpolated data with respect to some model; examples include statistical models (optimal interpolation and EOF methods) or least square projection on functional bases (OMA). In contrast, we propose an interpolation method that is based on the concept of extending data features (patterns) into missing regions. Thus, instead of directly interpolating the data or minimizing an interpolation error, the aim is to preserve coherent patterns that have evolved in the field measurements. Specifically, we propose a partial differential equation (PDE) based approach that is derived from a computer vision technique developed to restore missing information in digital images or videos; a process known as \emph{image inpainting}. While common inpainting techniques involve deconvolution and 2D filters on a frequency or wavelet domain, PDE-based methods form an important class of methods \citep{SCHONLIEB-2015}. A subclass of these methods aim to restore missing data by the {\em transport} of information from known to unknown regions. Transport can be achieved by {\em advective} or {\em diffusive} means. Both diffusive \citep{XU-2010} and advective \citep{BORNEMANN-2007, KORNPROBST-1997} information transport have been explored. Each method carriers limitations, e.g., diffusive transport may diminish important {\em features} while advective transport may produce artificial discontinuities and numerical challenges.

A more balanced approach to information transport is to construct anisotropic diffusion in a manner that preserves features \citep{PERONA-1990}. An example of this is the Cahn-Hillard equation (a non-linear generalization of diffusion equation), which  has been widely used to develop inpainting techniques \citep{BERTOZZI-2007, BURGER-2009, SCHONLIEB-2011}. 
Related to this approach, \cite{BERTALMIO-2000} combined advection and nonlinear diffusion to develop an inpainting technique that solves a PDE analogous to the Navier-Stokes equation. We extend this approach to the consideration of {\em field data}, and demonstrate the ability of this approach to effectively restore incomplete oceanographic data measurements. We focus on ocean surface current data measured by HF radar because this data is prone to incomplete coverage and restoring missing coverage is often essential to the effective utilization of the data. However, the method is extensible to other oceanographic quantities. 



\section{Method} \label{sec:method}


\subsection{Transport model} \label{sec:transport}

Let $\psi(\vect{x})$ 
denote a scalar field defined in $\Omega_d \subseteq \mathbb{R}^2$, which is the domain of interest. In image processing, the scalar field $\psi$ is often the image intensity (greyscale), or each of the color channels, and accepts integer values $ 0 \leq \psi \leq 255$. In our application $\psi: \Omega_d \to \mathbb{R}$ will represent either the east velocity $v_e(\vect{x},t)$ or north velocity $v_n(\vect{x},t)$ of the ocean surface current at a fixed time $t$. The fields $v_e(\vect{x})$ and $v_n(\vect{x})$ need not be assumed to satisfy any specific governing equations (e.g., Navier-Stokes or incompressibility) and hence can be processed separately. More generally, $\psi$ can represent any field variable of interest. 

\begin{figure}[t]
    \centering
    \footnotesize{
    \includegraphics[width=0.37\textwidth]{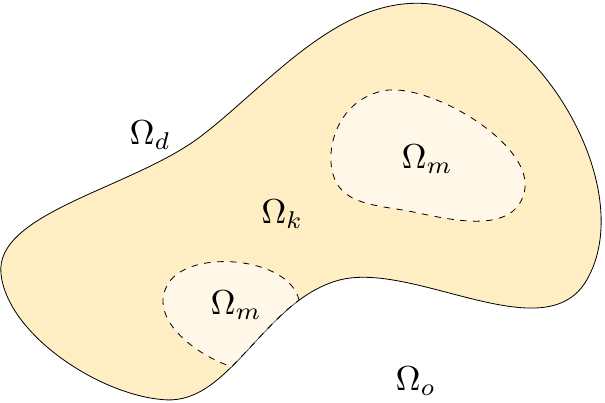}
    }
    \caption{Dataset domain $\Omega_d \subseteq \mathbb{R}^2$ is decomposed into disjoint sets $\Omega_d = \Omega_k \displaystyle{\cup} \Omega_m$ where in $\Omega_k$ the data field $\psi(\vect{x})$ is known and in $\Omega_m$ the data field is missing. $\Omega_o$ represents the region outside the data domain.}
    \label{fig:domain-decompose}
\end{figure}

Let $\Omega_o$ be all space outside the domain of interest, so that $\mathbb{R}^2 = \Omega_d \cup \Omega_o$ is a disjoint union. Decompose the domain $\Omega_d = \Omega_k \cup \Omega_m$ so that in $\Omega_k$ the function $\psi$ is known and in $\Omega_m$ the value of $\psi$ is missing (see Figure \ref{fig:domain-decompose}). The goal is to determine $\psi$ in $\Omega_m$ so that the overall field in $\Omega_d$ is {\em consistent}. By consistent we mean $\psi$ is second order continuous on $\partial \Omega_m$, and, roughly speaking, the patterns of $\psi$ around $\Omega_m$ continue inside the missing domain. The later is based more on qualitative assessment (although the restoration error will be rigorously quantified). To accomplish this, we aim to advect data {\em features} in the neighborhood of $\Omega_m$ toward the missing areas. In the following we explain the construction of a PDE that diffuses and advects the features of the scalar field $\psi$.

The Laplacian $\omega \coloneqq -\nabla^2 \psi$ is commonly used to identify features in 2D field data. In image processing $-\omega$ is often referred to the \emph{image smoothness} since it represents the curvature of the scalar field $\psi$.  We use $\omega$ as the desired quantity to be advected from a neighborhood around $\partial \Omega_m$ into the missing domain $\Omega_m$. To maintain features, $\omega$ is advected along the levelsets of $\psi$. 

Define $\vect{u} \coloneqq \nabla^{\perp} \psi$ to be the vector orthogonal to the gradient $\nabla \psi$, \ie $\nabla^{\perp} \psi = (\partial \psi / \partial y) \vect{i} - (\partial \psi / \partial x) \vect{j}$. Flow that is induced by the vector field $\vect{u}$ traces the levelsets of $\psi$. Therefore, to advect and preserve $\omega$ along the levelsets of $\psi$, the material derivative of $\omega$ along the flow of $\vect{u}$ should vanish, i.e.,
\begin{equation}
    \frac{\mathrm{D}\omega}{\mathrm{d}t} = \frac{\partial \omega}{\partial t} + \vect{u} \cdot \nabla \omega = 0 .\label{eq:adv}
\end{equation}
We note that time $t \in \mathbb{R}^{+}$ here is not the actual time in arguments of $v_e(\vect{x},t)$ or $v_n(\vect{x},t)$. Rather, here $t$ is a strictly increasing variable that is used for the forward advection of $\omega$ for the data restoration process. The desired propagation of information of $\psi$ from $\Omega_k$ to $\Omega_m$ is achieved when the solution of \eqref{eq:adv} converges to a steady state solution for some sufficiently large $t$. The steady state solution of \eqref{eq:adv} satisfies $\vect{u} \perp \nabla \omega$.


To stabilize the pure advection equation \eqref{eq:adv} as well as to more smoothly propagate the information of $\psi$ to missing regions, a nonlinear \emph{anisotropic diffusion} term is added to \eqref{eq:adv} by
\begin{equation}
    \frac{\partial \omega}{\partial t} + \vect{u} \cdot \nabla \omega = \nu \nabla \cdot \left( g \nabla \omega \right) . \label{eq:adv-diff}
\end{equation}
where $g = g(| \nabla \omega |)$ 
is a nonlinear diffusivity and $\nu \in \mathbb{R}^{+}$ is a weight parameter. Occasionally $g = g(|\nabla \psi|)$ is used \citep{FISHELOV-2006}, however, in this work we use $g = g(| \nabla \omega |)$.

Note that \eqref{eq:adv-diff} is analogous to the 2D incompressible vorticity transport equation in fluid mechanics. Namely, the image intensity $\psi$ would be the stream-function, $\vect{u}$ would be the fluid velocity, smoothness $\omega$ would be vorticity, and $\nu  g$ would be (nonlinear) fluid viscosity. Constant diffusivity ($g = 1$) recovers the classical vorticity transport equation for a Newtonian fluid. Note, despite this analogy, there is no implied relationship between $\vect{u}$ and the east or north velocity fields $v_e(\vect{x},t)$ or $v_n(\vect{x},t)$ describing the original data.

In the absence of viscosity, \ie $\nu g = 0$, features defined by $\omega$ are purely advected by $\vect{u}$. At the other extreme, when $\nu g \gg 1$, the diffusion term is dominant, and the PDE acts as smoothing filter that blurs the data locally with minimal advection of features.  

In practice, a monotonically decreasing function is used for viscosity so that $g(0) = 1$ and $g(\infty) = 0$. Such functional form produces a directional diffusion that aims to preserve data features based on how well defined these features are \citep{BLACK-1998}. A common choice is a Perona-Malik anisotropic diffusion \citep{PERONA-1990} given by $g(s) = (1+s^2)^{-1}$ or $g(s) = e^{-s^2}$ where the generic argument $s$ is a scalar field defined depending on the context. Namely for anisotropic diffusion of vorticity $s \coloneqq | \nabla \omega |$ \citep{BERTALMIO-2001} and thus  
\begin{equation}
    g(| \nabla \omega |) = \left[ 1 + \left( \frac{| \nabla \omega |}{K} \right)^2 \right]^{-1}, \label{eq:g}
\end{equation}
where $K > 0$ is used to non-dimensionalize $g$. The above form enhances diffusion in areas where the magnitude of vorticity gradient is low, and diminishes the diffusion where the magnitude of vorticity gradient is high. The addition of the nonlinear, anisotropic diffusion enhances the numerical stability and convergence of solving \eqref{eq:adv-diff}, while preserving features. 

Since \eqref{eq:adv-diff} is an equation for $\omega$ and not the original field $\psi$ that we seek to restore, \eqref{eq:adv-diff} is coupled with the Poisson equation
\begin{equation}
    \omega + \nabla^2 \psi = 0. \label{eq:Poisson}
\end{equation}
The set of two second order coupled PDEs (\ref{eq:adv-diff} and \ref{eq:Poisson}) are solved with initial and boundary conditions described below.




\subsection{Initial and boundary conditions} \label{sec:bc}

The local transport of information can be controlled by modifying how ``boundary conditions'' for the above PDEs are applied. Traditionally, the solution of \eqref{eq:adv-diff} in $\Omega_m$ would be informed from boundary conditions that specify the value of $\psi$ (or its normal derivative) on $\partial \Omega_m$. However, we seek to inform the solution of \eqref{eq:adv-diff} in $\Omega_m$ using the value of $\psi$ over a local neighborhood around $\Omega_m$, as opposed to just on the boundary $\partial \Omega_m$. 

\begin{figure}[h]
    \centering
    \footnotesize{
    \includegraphics[width=0.45\textwidth]{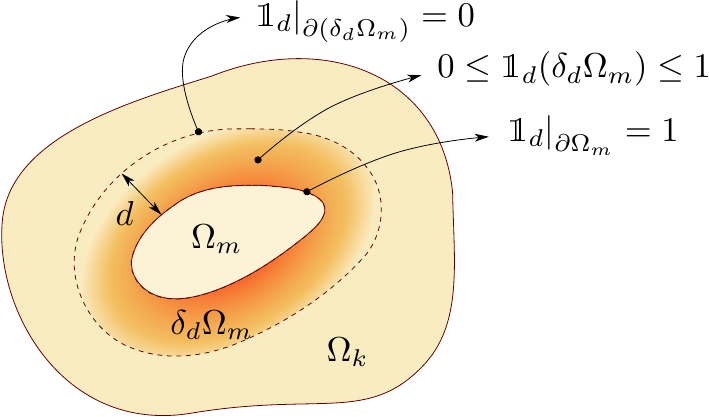}
    }
    \caption{Boundary conditions needed to solve \eqref{eq:adv-diff} in $\Omega_m$ are effectively generated by gradually extending the transport equation \eqref{eq:adv-diff} beyond $\Omega_m$ into an inflated region $\delta_{d} \Omega_m$. In this way, information over $\delta_{d} \Omega_m$, as opposed to $\partial \Omega_m$, is used to inform the solution of \eqref{eq:adv-diff} in $\Omega_m$. The smooth step function $\mathbbm{1}_{d}$ gradually diminishes use of information away from $\partial \Omega_m$.}
    \label{fig:BoundaryCondition}
\end{figure}

Define a boundary band $\delta_{d} \Omega_m$ as an inflation of $\Omega_m$ by a small distance $d > 0$, as shown in Figure \ref{fig:BoundaryCondition}. Over this band, the PDE \eqref{eq:adv-diff} is modified as
\begin{equation}
    \frac{\partial \omega}{\partial t} = \mathbbm{1}_{d}(\vect{x}) \mathcal{L}(\vect{u},\omega,t) \label{eq:adv-diff-alt},
\end{equation}
where the operator $\mathcal{L}$ is
\begin{equation*}
    \mathcal{L}(\vect{u},\omega,t) \coloneqq - \vect{u} \cdot \nabla \omega + \nu \nabla \cdot \left( g \nabla \omega \right).
\end{equation*}
The function $\mathbbm{1}_{d} \in C^2(\Omega_d)$ is a smooth variant of an indicator function such that 
\begin{equation*}
    \mathbbm{1}_{d}(\vect{x}) = 
    \begin{cases}
        1, & \vect{x} \in \Omega_m, \\
        0, & \vect{x} \in \Omega_k \setminus \left( \Omega_m \cup \delta_{d} \Omega_m \right),
    \end{cases}
\end{equation*}
and continuously varies in the range $0 \leq \mathbbm{1}_{d}(\vect{x}) \leq 1$ over the band $\delta_{d} \Omega_m$. We use a cubic Hermite spline $\mathbbm{1}_d(r) = 2r^3-3r^2+1$ in terms of outward distance $0 \leq r \leq d$ from $\partial \Omega_m$. 

Although $\psi$ is known in $\delta_d \Omega_m$, we treat $\psi$ to be unknown in $\delta_d \Omega_m$, and solve the system of equations \eqref{eq:adv-diff-alt} and \eqref{eq:Poisson} in $\Omega_m \cup \delta_d \Omega_m$ rather than only $\Omega_m$. The initial condition of $\psi$ in $\delta_d \Omega_m$ is set to the known measurements, and the initial condition for $\psi$ in $\Omega_m$ is set to the local average of $\psi$ in $\delta_d \Omega_m$. Based on this, the initial condition for $\omega = -\nabla^2 \psi$ can be computed in $\Omega_m \cup \delta_d \Omega_m$. Since the initial condition for $\psi$ is constant in $\Omega_m$, the initial condition of $\omega$ vanishes in the interior of $\Omega_m$ but is non-zero near the boundary $\partial \Omega_m$ and inside $\delta_d \Omega_m$.

A Dirichlet boundary condition is applied for both $\psi$ and $\omega$ on the inflated boundary $\partial (\delta_d \Omega_m)$. The value of  $\psi$ is directly set based on the known measurements. To maintain a consistent boundary condition for $\omega$, at each numerical step we compute $\omega = -\nabla^2 \psi$ on $\partial (\delta_d \Omega_m)$ to update the Dirichlet boundary condition for $\omega$. In the special case where the $\partial \Omega_m$ overlaps the boundary of $\Omega_d$ (see for example Figure \ref{fig:domain-decompose}), a homogeneous Neumann boundary condition is applied. Numerical integration proceeds until convergence of the solution to a steady-state. Once the solution converges, the values of $\psi$ in $\delta_d \Omega_m$ are replaced with their known values.


\subsection{Numerical considerations} \label{sec:implementation}

Ocean surface velocity datasets are usually provided on structured geographic grid (longitudes and latitudes). The scale of typical HF radar coverage are often small enough so that the geographic grid of longitudes and latitudes can be directly used without projecting geographic grid. The differential equations can be discretized with a finite difference scheme on a 2D Cartesian coordinate system. For large domains or data that are provided on unstructured grids, other numerical schemes such as the finite element method are more convenient.

To improve convergence, \eqref{eq:Poisson} was replaced with the pseudo-steady parabolic equation \citep{BERTALMIO-2000}
\begin{equation}
    \frac{\partial \psi}{\partial t} = \alpha \left( \nabla^2 \psi + \omega \right) \label{eq:poisson-relaxed}
\end{equation}
where $\alpha$ is a relaxation parameter. The steady-state solution of \eqref{eq:poisson-relaxed} converges to the Poisson equation \eqref{eq:Poisson}. Spatial derivatives in \eqref{eq:poisson-relaxed} and \eqref{eq:adv-diff-alt} were evaluated using second-order accurate central differences, and time stepping of each as was performed using implicit Euler. 



Low viscosity affects numerical stability. To this end, we set $\nu$ such that the grid cell Reynolds number
\begin{equation*}
    \mathrm{Re} = \frac{|\vect{u}| \Delta x}{\nu} \lessapprox 1,
\end{equation*}
holds. Thus, in the vorticity transport equation, the diffusion becomes the dominant term, and the stability is carried by the Fourier diffusion number,
\begin{equation*}
    \mathrm{Fo}_{d} = \frac{\nu \Delta t}{\Delta x^2},
\end{equation*}
which indicates the balance between the diffusion term and the unsteady term. We scale the spatial dimensions such that $\Delta x = \Delta y = 1$, and set $\Delta t$ such that $\mathrm{Fo}_d$ remains small. In particular, we demonstrate in the results below that  $\mathrm{Fo}_d \approx 0.2$ yields best convergence.  Similarly $K$ is chosen to facilitate convergence and results below demonstrate that $K>0.1$ is suitable. The last model parameter is the dilation radius $d$ of boundary $\partial \Omega_m$. This will depend on the relative size of the missing patch of data, as further discussed below. We also demonstrate the solution is generally not sensitive to the choice of $d$.  

%



\section{Application to Oceanographic Data} \label{sec:application}

We apply the method of \S \ref{sec:method} to the HF radar measurements of ocean surface currents along the coast of Martha's Vineyard and northern California. These two domains have different spatial coverage scales and differing degree of missing data; thus each offers a unique application for testing. 


\subsection{Martha's Vineyard HF radar data} \label{sec:MVCO}

The first test case is data from the HF radar network at Martha's Vineyard Coastal Observatory (MVCO) operated by Woods Hole Oceanographic Institution (WHOI), Massachusetts. The MVCO is a unique high precision system of HF radars with \SI{400}{\m} resolution that covers $20 \times 20$ \si{\kilo\m} area south Martha's Vineyard island. Details on the MVCO HF radar can be found in, e.g., \citep{KIRINCICH-2012, KIRINCICH-2016}. Based on its robust local coverage, we use this data primarily for validation.


\subsubsection{Validation testing} \label{sec:validation}

\begin{figure}[h]
    \centering
    \footnotesize{
    \includegraphics[width=0.5\textwidth]{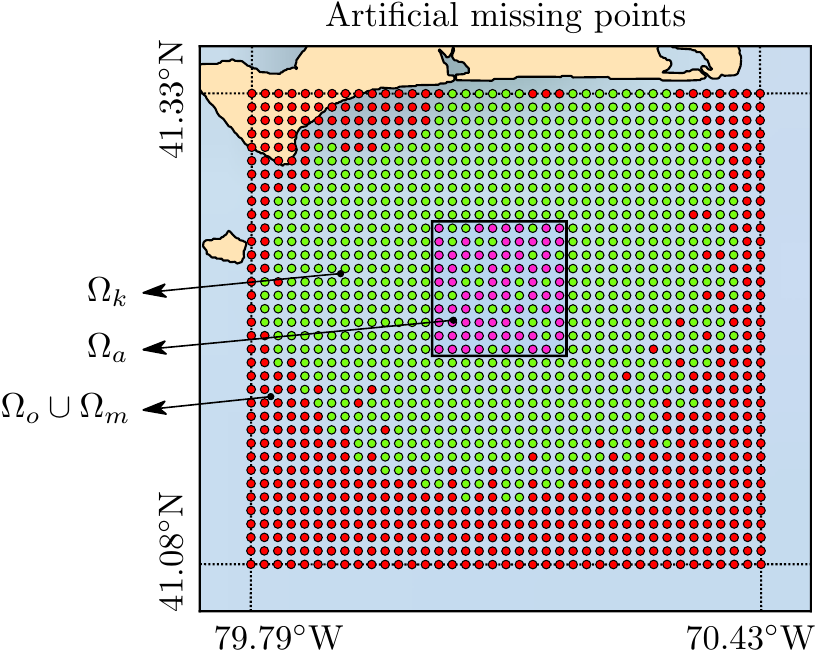}
    }
    \caption{Domain for MVCO data. The green and pink points compose $\Omega_k$, where velocity measurements are known. The pink points are excluded for testing, and the percentage and locations of these missing points is varied. Red points indicate missing measurements.}
    \label{fig:MVCO_locate_art}
\end{figure}

Figure \ref{fig:MVCO_locate_art} shows the location of the original geographic grid of MVCO data. The north side of the figure is Martha's Vineyard island. The green and pink dots represent $\Omega_k$ where the values of velocities are known. The surrounding red dots represent unknown data, and can be considered either in the missing domain $\Omega_m$ or the outside domain $\Omega_o$. We will distinguish between $\Omega_m$ and $\Omega_o$ later in \S \ref{sec:MVCO-app}. Inside the black square of $10 \times 10$ points, we synthetically introduce $N$ missing points shown in pink and denoted by $\Omega_a \subset \Omega_k$. Since the true values of velocity on $\Omega_a$ are known, they can be used to test and validate the method.

To compare the restored results, define the normalized root-mean-square error (NRMSE) of a function $q: \Omega_d \to \mathbb{R}$ by
\begin{equation}
    \epsilon(q,\Omega_a,N) = \sqrt{ \frac{1}{N} \sum_{\vect{x}_i \in \Omega_a} \left(\frac{q_r(\vect{x}_i) - q(\vect{x}_i)}{q(\vect{x}_i)} \right)^2 }, \label{eq:NRMSE}
\end{equation}
where $q$ is the true value and $q_r$ is the restored value. Recall that we apply the restoration algorithm separately to the east $\psi=v_e$ and north $\psi=v_n$ velocity field. However, we are practically more concerned with the velocity vector magnitude and direction. Therefore, in \eqref{eq:NRMSE}, we consider $q$ to be either the velocity magnitude $q=|\vect{v}| = \sqrt{v_e^2 + v_n^2}$ or the velocity direction $q = \arg(\vect{v}) = \tan^{-1}(v_n / v_e)$. We often refer to the NRMSE as the ``restoration error''. 

To achieve a comprehensive statistical comparison, several realizations of excluded measurements are considered. Namely, we varied the number of missing points $N$, and for each value of $N$ we performed $M=100$ realizations of randomly excluding $N$ data points inside the $10 \times 10$ square. This enables the errors $\epsilon(q, \Omega_a, N)$ to then be averaged over $M$ to compute its mean and standard deviation.

\begin{figure}[h]
    \centering
    \footnotesize{
    \includegraphics[width=\textwidth]{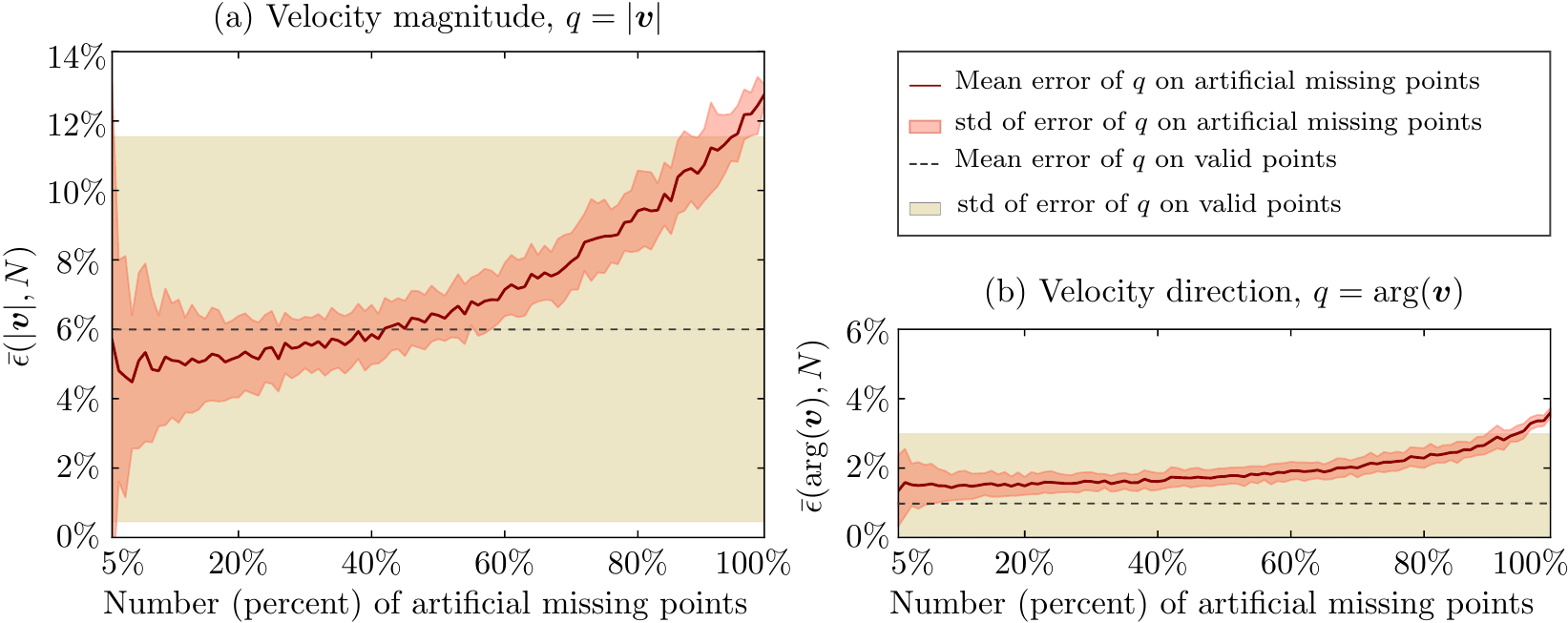}
    }
    \caption{Reconstruction error $\bar{\epsilon}$ and standard deviation (std) as a function of missing data percentage $N$. (a) Error in velocity magnitude, $q = |\vect{v}|$. (b) Error in velocity direction, $q = \arg(\vect{v})$.}
    \label{fig:error_missingpoints}
\end{figure}

The domain $\Omega_a$ is restored for July 1\textsuperscript{st}, 2014, 10:00 UTC with an inflated boundary band size $d = 5$ points counted from the missing domain. Figure \ref{fig:error_missingpoints} represents the NRMSE of the restored velocity data by varying the number of missing points $N$ in terms of percentage relative to points in the square domain. Panels (a) and (b) of the figure correspond to the NRMSE of velocity magnitude $\epsilon(|\vect{v}|,N)$ and direction $\epsilon(\arg(\vect{v}),N)$ respectively. The solid lines are the mean $\bar{\epsilon}$ over all $M$ realizations, and the shaded neighborhood is the standard deviation.

In addition, the east velocity error $e_{v_e}(\vect{x}) = \delta v_e(\vect{x})$ and north velocity error $e_{v_n}(\vect{x}) = \delta v_n(\vect{x})$ of HF radar measurements are often available or can be estimated from the Geometric Dilution of Precision (GDOP) \citep{CHAPMAN-1997,LIPA-2003}. The dashed lines and the shaded bounds in Figures \ref{fig:error_missingpoints} (a) and (b) are the mean and standard deviation of the velocity magnitude error $\delta |\vect{v}| = (v_e \delta v_e + v_n \delta v_n) / |\vect{v}|$ and the velocity direction error $\delta \arg(\vect{v}) = (1 + \arg(\vect{v})^2)^{-1} (v_e \delta v_n - v_n \delta v_e) / v_e^2$ of HF radar measurements, respectively.


It can be observed from Figure \ref{fig:error_missingpoints} that the restoration error in velocity magnitude $\epsilon(|\vect{v}|,N)$ and direction $\epsilon(\arg(\vect{v}),N)$ are both within the same order of magnitude of the intrinsic measurement errors $\delta |\vect{v}|$ and $\delta \arg(\vect{v})$ for up to around 95\% missing data. Note that the error in velocity direction is nearly an order of magnitude smaller than the error in velocity magnitude. Thus, streamline patterns, which depend on velocity direction, can be restored with higher accuracy. 

\begin{figure}[!htpb]
    \centering
    \footnotesize{
    \includegraphics[width=\textwidth]{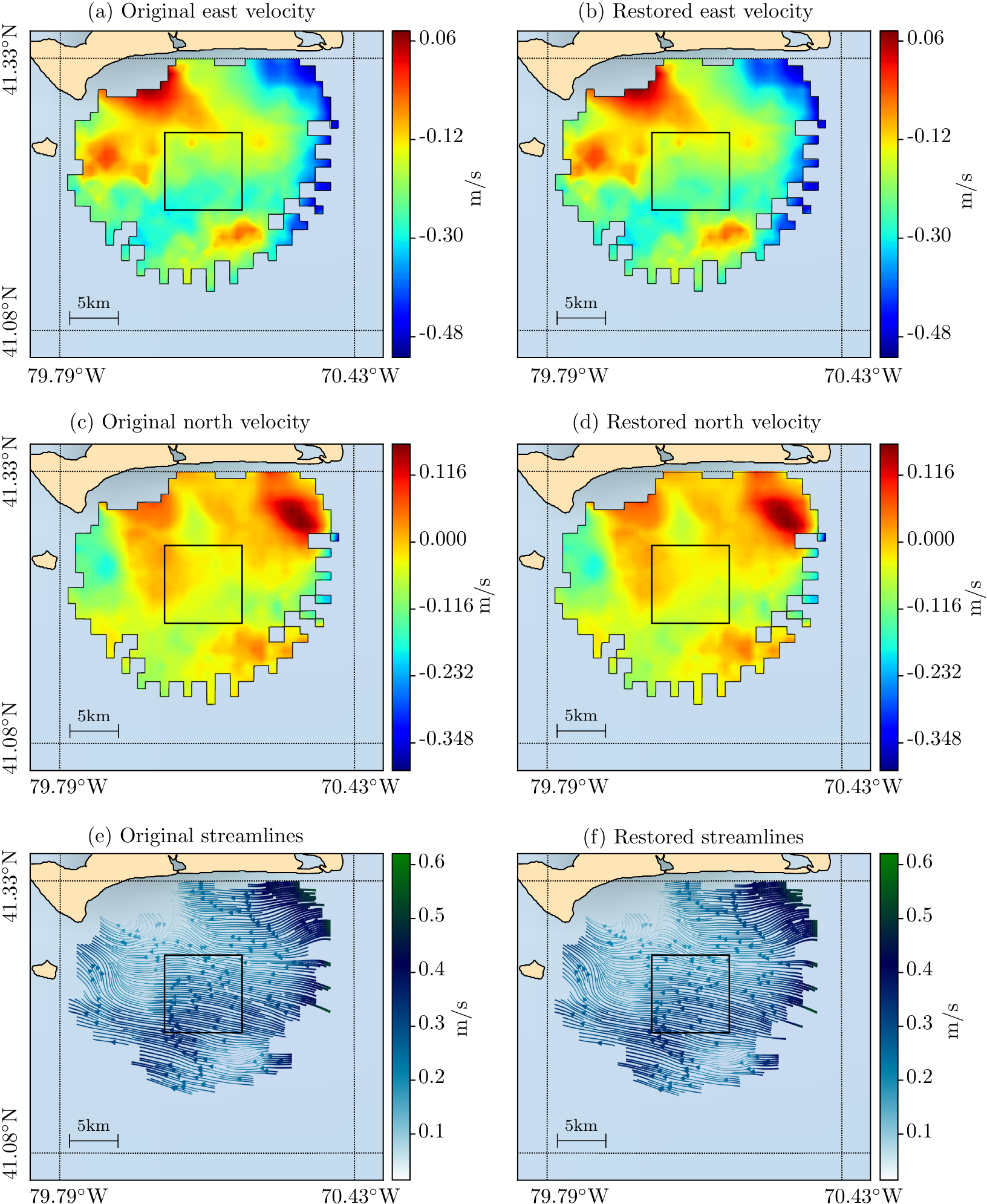}
    }
    \caption{Martha's Vineyard HF radar velocity data are shown on July 1\textsuperscript{st}, 2014 at 10:00 UTC. Original (left column) and restored data (right column) are compared for the synthetic missing domain $\Omega_a$ inside the black square for $|\Omega_a| = N = 70$ missing points. The third row is the resultant streamlines colored by velocity magnitude.}
    \label{fig:MVCO_velocities_artificial}
\end{figure}

Because NRMSE represents a spatially-averaged error, it is useful to visually compare the restored field with the true field. The instance of $N = 70$ missing points of the Figure \ref{fig:MVCO_locate_art} on July 1\textsuperscript{st} 2014, 10:00 UTC is shown in Figure \ref{fig:MVCO_velocities_artificial},  where the original and restored velocities and their streamlines are compared. Despite the fact that $70$ percent of the area in the black square is originally missing, the true and restored fields are visually indistinguishable.

\subsubsection{Boundary band size} \label{sec:boundary_size}

\begin{figure}[!htpb]
    \centering
    \footnotesize{
    \includegraphics[width=\textwidth]{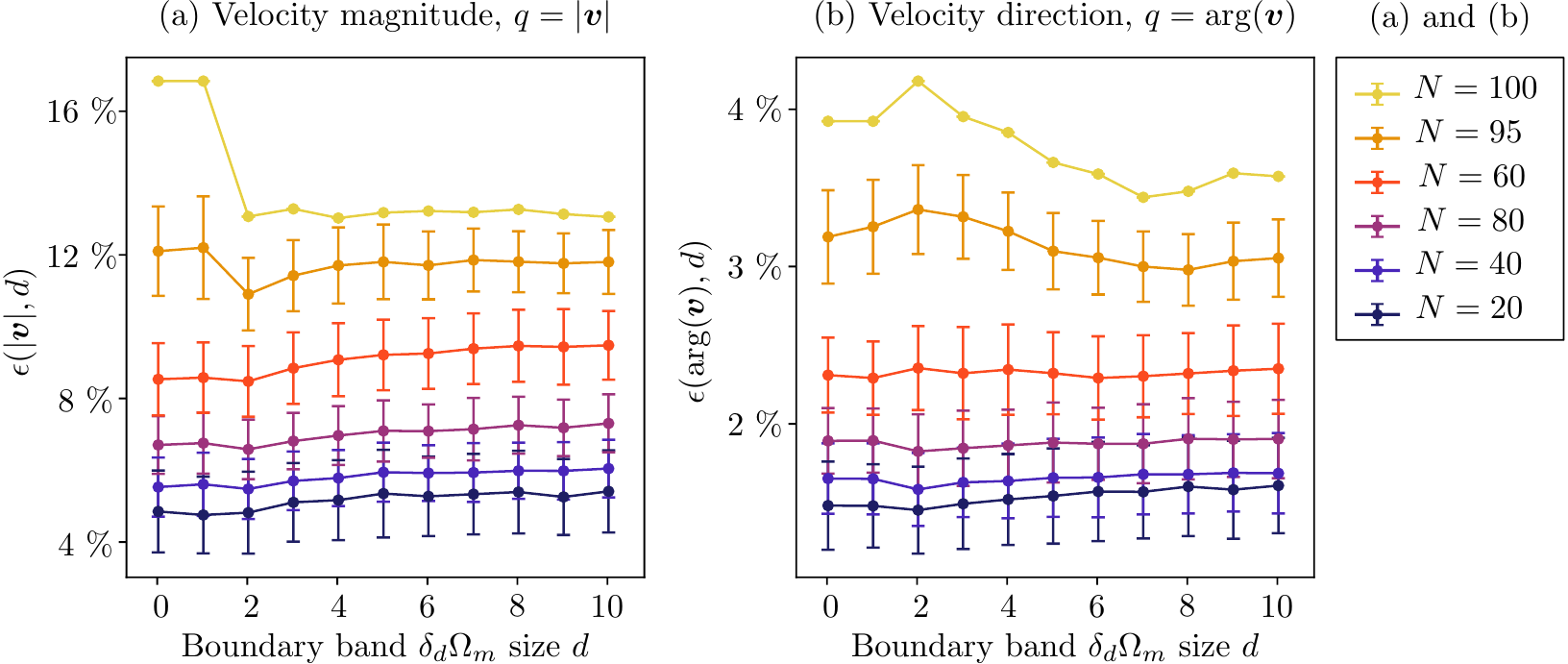}
    }
    \caption{The NRMSE of comparing the synthetic missing points and the true velocities by varying the size $d$ of the boundary band $\delta_d \Omega_m$ and the number of synthetic missing points $N$. Size $d$ is measured by the grid points from $\partial \Omega_m$. Figure (a) corresponds to the error of velocity magnitude and (b) corresponds to the error of velocity direction.}
    \label{fig:error_varying_d}
\end{figure}

The effect of boundary size $d$ on the accuracy of the restoration is shown in Figure \ref{fig:error_varying_d}, where again panels (a) and (b) corresponding to restoration error in velocity magnitude and direction, respectively. The size $d$ is counted as number of layers of data points from the missing domain. Several choices for $N$ were tested, and $M = 100$ realizations for each choice of $N$ were considered. The solid lines are the average restoration error, and the bars denote standard deviations. The results in Figure \ref{fig:error_varying_d} indicate that the restoration is relatively insensitive to the choice of $d$, and the $d$ covering a few layers of data is acceptable. As shown, covering at least a few layers appears most important when the amount of missing data is large. 

\begin{figure}[!htpb]
    \centering
    \footnotesize{
    \includegraphics[width=\textwidth]{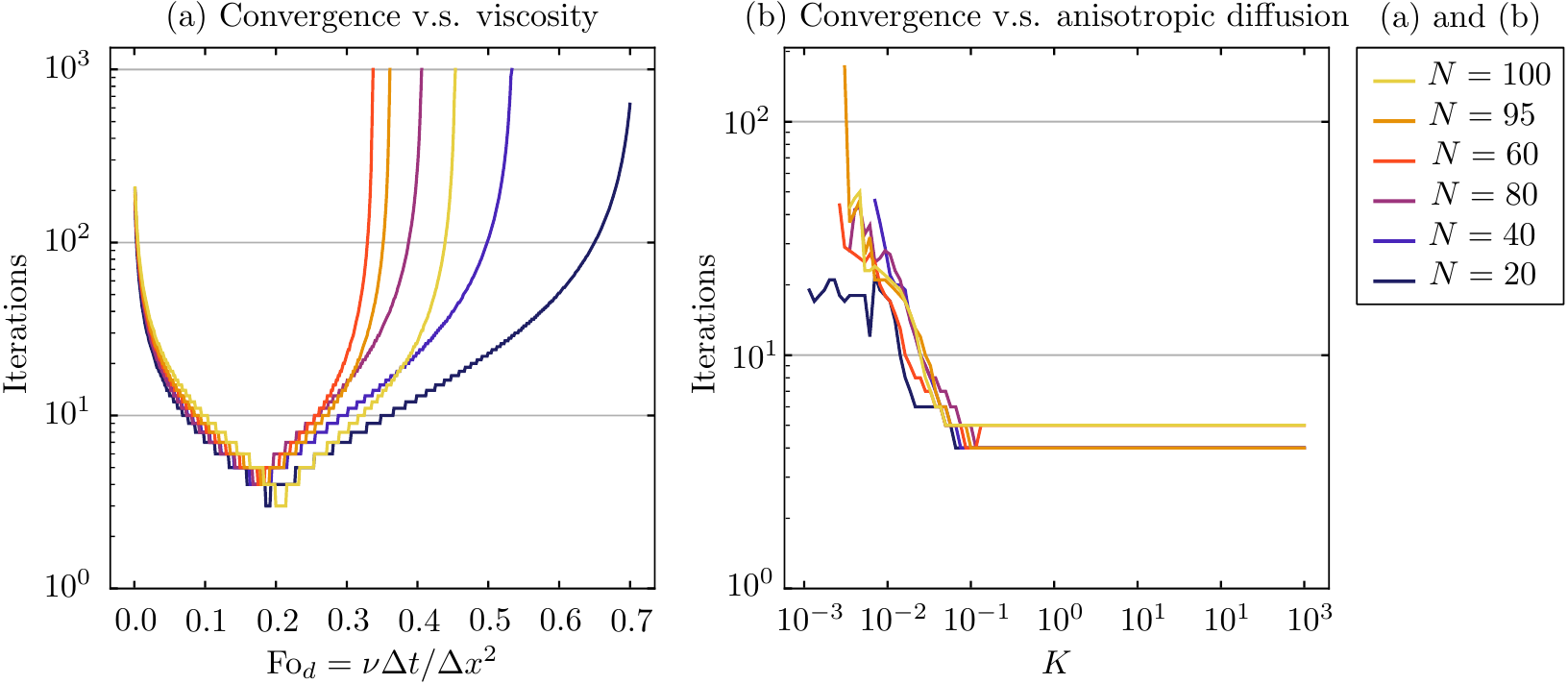}
    }
    \caption{The effect of model parameters on solution convergence for east velocity data. The convergence is measured by number of iterations to achieve the relative error $10^{-3}$. Results are presented for varying percent $N$ of missing data in Figure \ref{fig:MVCO_locate_art}. (a) Viscosity $\nu$ is varied as the dimensionless Fourier diffusion number $Fo_d = \nu \Delta t / \Delta x^2$ while $K = 10^3$ is fixed. (b) $K$ is varied while $\nu = 20$ and $\Delta t = 0.01$ are fixed.}
    \label{fig:error_varying_K_nu}
\end{figure}

\subsubsection{Parameter selection} \label{sec:parameters}

The PDE transport approach has few parameters. Parameters $\nu$ and $K$ were observed to have no significant effect on the accuracy on the results, however, they influence the convergence rate of the numerical procedure. To test convergence, the  $L_{\infty}$ norm of the relative error in $\psi$ over successive iterations was considered. Figure \ref{fig:error_varying_K_nu} shows the number of iterations taken to achieve a relative error less than $10^{-3}$ for multiple choices of $N$ (same domain and timeframe as in Figure \ref{fig:MVCO_locate_art}). Figure \ref{fig:error_varying_K_nu} (a) considers the effect of viscosity in terms of dimensionless Fourier number $\mathrm{Fo}_d$. $\mathrm{Fo}_d \approx 0.2$ yields best convergence over a broad range of $N$. Practical choices are $\Delta t = 0.01$ and $5 \leq \nu \leq 20$. Such viscosity maintains the Reynolds number in a reasonable range to employ both convective and diffusive terms. Figure \ref{fig:error_varying_K_nu} demonstrates that best convergence is achieved for any $K > 10^{-1}$. The convergence rates shown in Figure \ref{fig:error_varying_K_nu} for parameters $K$ and $\nu$ were found to be more or less similar over various datasets we have considered. The run-time for a typical application, such as the tests above, with appropriately-selected parameters was around a few seconds on a single core processor.


\begin{figure}[h]
    \centering
    \footnotesize{
    \includegraphics[width=0.85\textwidth]{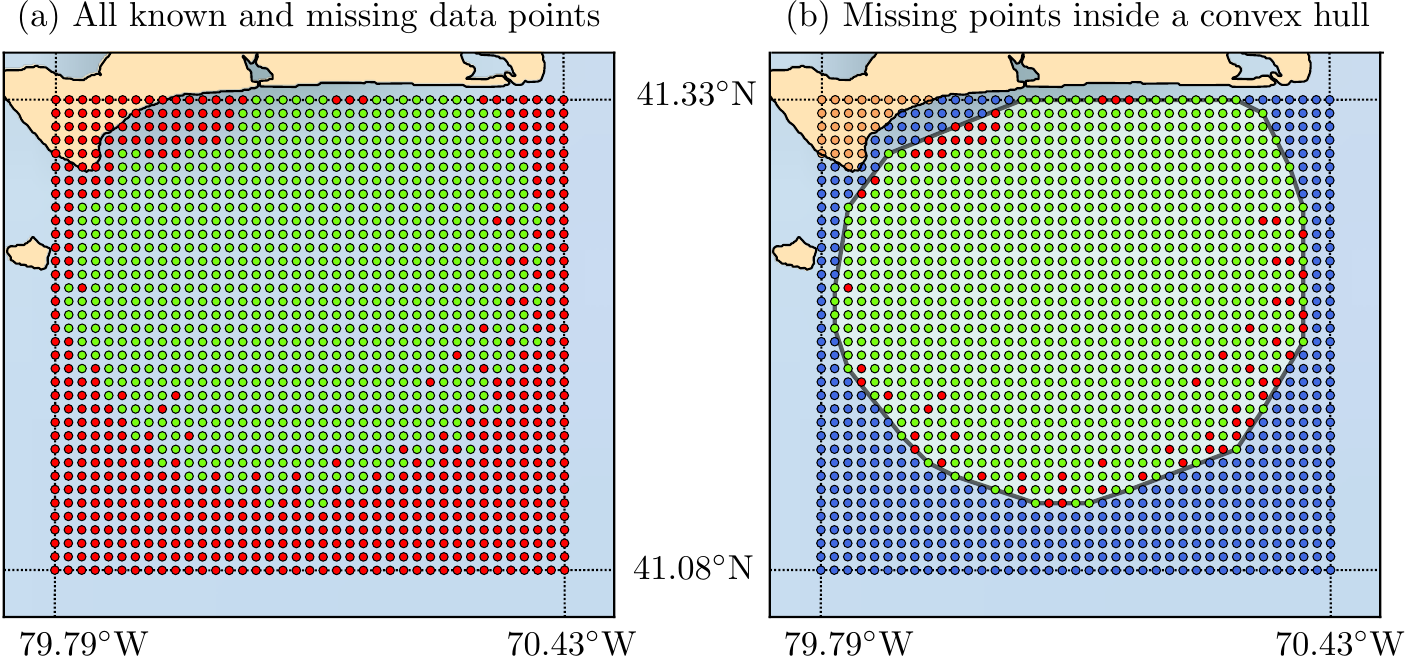}
    }
    \caption{Geographic location of data grid for Martha's Vineyard HF radar. Known domain $\Omega_k$  are shown with green points. (a) All unknown points on the whole grid are shown in red which are $\Omega_m \cup \Omega_o$. (b) Missing points $\Omega_m$ are separated from outside domain $\Omega_o$ with a convex hull around $\Omega_k$.}
    \label{fig:missingpoints}
\end{figure}
\begin{figure}[!htpb]
    \centering
    \footnotesize{
    \includegraphics[width=\textwidth]{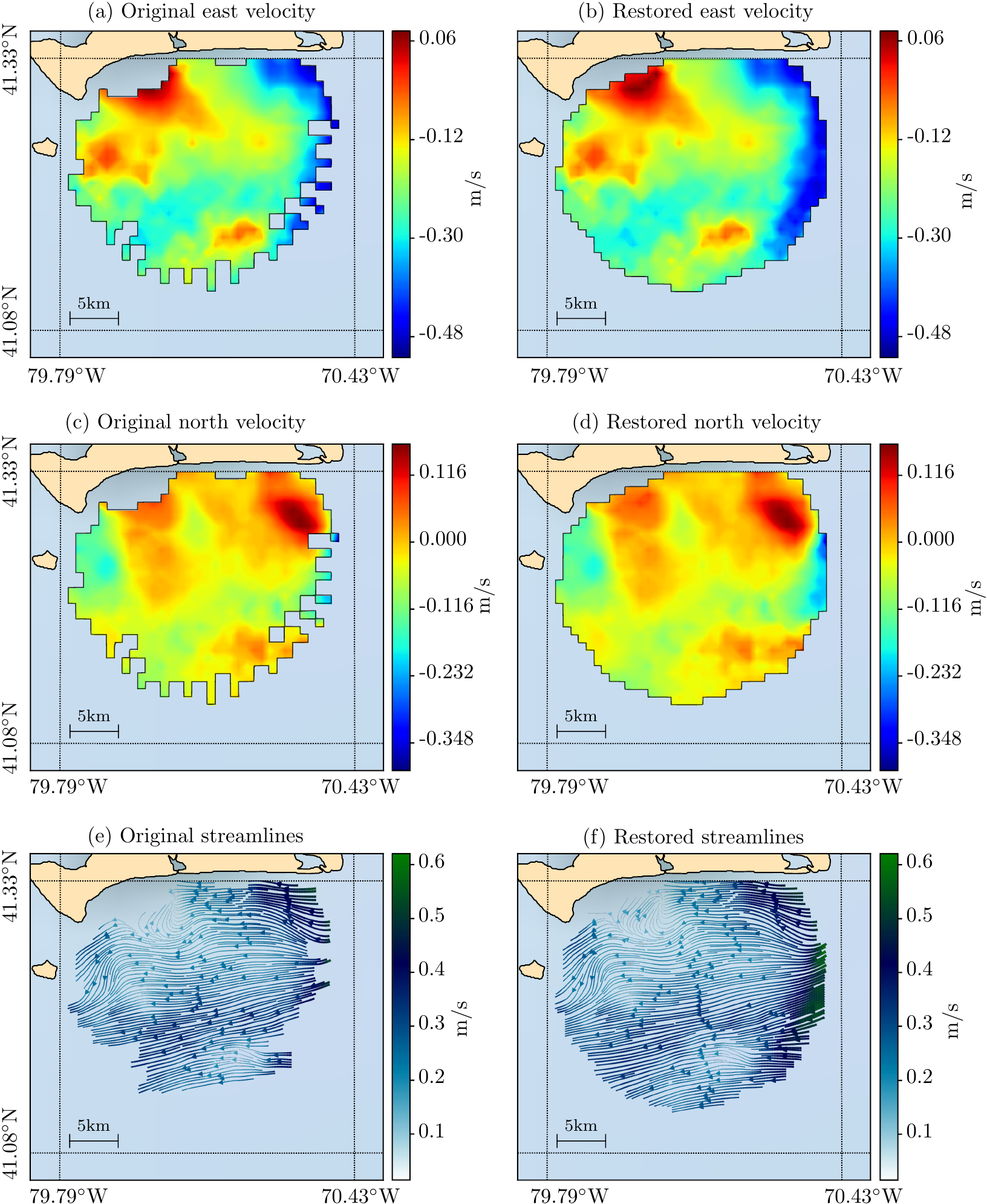}
    }
    \caption{The velocity fields of Figure \ref{fig:MVCO_velocities_artificial} are fully restored inside the convex hull around $\Omega_k$. Left and right columns are the original and restored data respectively. First and second rows are east and north velocity fields respectively. The third row is the resultant streamlines of each dataset colored by velocity magnitude. For the purpose of visualization, the continuous fields are obtained by linear interpolation.}
    \label{fig:MVCO_velocity_restored}
\end{figure}

\subsubsection{Practical application} \label{sec:MVCO-app}

We aim to restore {\em actual} missing measurements inside and on the boundary of the Martha's Vineyard dataset considered above. Figure \ref{fig:missingpoints} (a) displays known data in green and unknown data in red. The boundary of the data domain $\Omega_d = \Omega_k \cup \Omega_m$ is not well-defined. We distinguish $\Omega_m$ from $\Omega_o$ by a convex hull that encloses $\Omega_k$ as shown in Figure \ref{fig:missingpoints} (b). Similar to the previous section, a dilation boundary of $d = 5$ was considered. The restored data is compared with the original data in Figure \ref{fig:MVCO_velocity_restored}. The continuity of data patterns that are extended to the missing domain can be readily observed on velocity fields as well as the streamlines.


\subsection{Northern California HF radar data} \label{sec:CA}

In this section we apply the restoration method coastal measurement form northern California available from the National HFRADAR Network Gateway at the Coastal Observing Research and Development Center (CORDC). The coastal ocean circulation of northern California has been broadly studied, for instance in Monterey bay \citep{PADUAN-1996, PADUAN-2004, SHULMAN-2009}, Bodega bay \citep{KAPLAN-2005} and subtidal velocity circulations \citep{LARGIER-1993, DEVER-1997}. We consider a \SI{200}{\kilo\m} region along the northern California coast extending around \SI{100}{\kilo\m} offshore. Data coverage for this region is more sporadic than for the MVCO data considered above, and better represents meso-beta coverage versus meso-gamma coverage. 


\subsubsection{Detecting the domain boundary}

\begin{figure}[h]
    \centering
    \footnotesize{
    \includegraphics[width=0.85\textwidth]{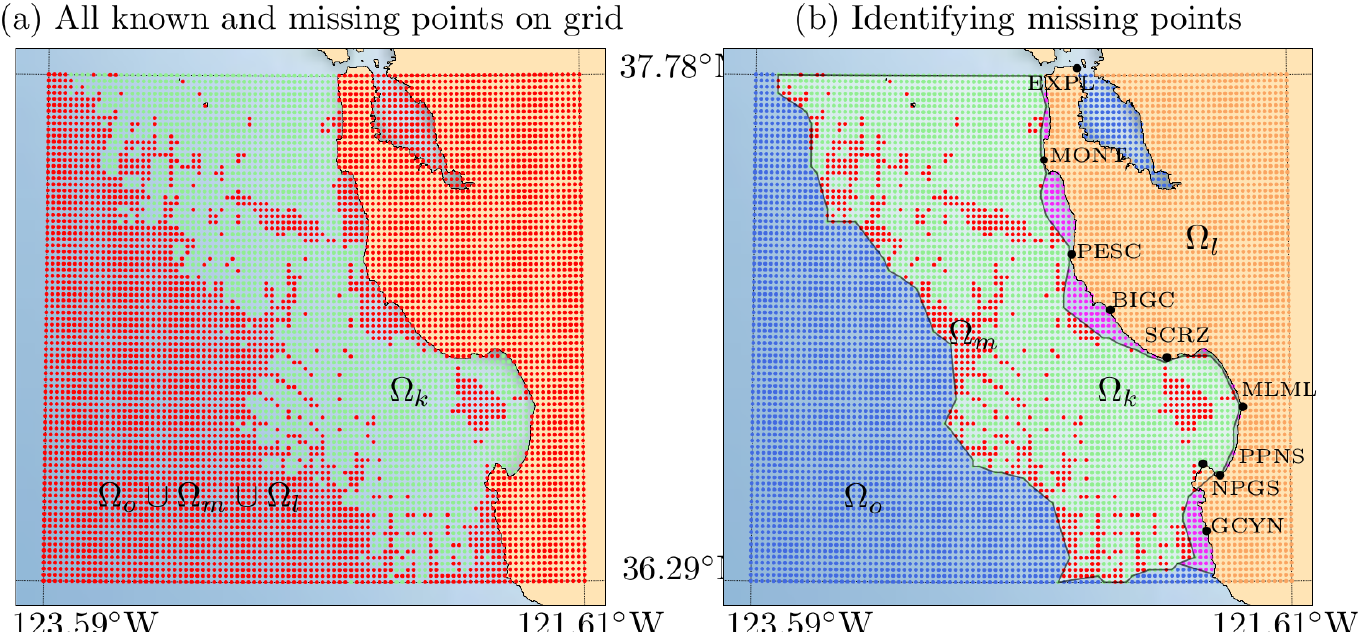}
    }
    \caption{Green points represent known measurements. (a) Red points are considered outside the domain $\Omega_o$, missing $\Omega_m$, or on land $\Omega_l$. (b) We distinguish points outside the domain $\Omega_o$ (blue) from those missing $\Omega_m$ (red) using an $\alpha$-shape around $\Omega_k$. Pink points represent extension of missing data $\Omega_m$ to the coastline (see \S \ref{sec:extend-coast}). HF radar sites are indicated along with code names.}
    \label{fig:MontereyBay_LocatingMissingPoints}
\end{figure}

Figure \ref{fig:MontereyBay_LocatingMissingPoints} (a) shows the availability of measurements mapped to a rectangular grid of \SI{2}{\kilo\m} resolution. The green points denote the domain of known measurements $\Omega_k$. The red points represent  missing data inside the domain (denoted $\Omega_m$) and outside the domain (denoted $\Omega_o$). $\Omega_m$ and $\Omega_o$ should be formally distinguished for proper assignment of boundary conditions in the restoration process.  Applying a convex hull around $\Omega_k$ does not produce a compact boundary for $\Omega_k$, and leads to unnecessary exclusions of concave regions. For geometries such as these, the boundary of $\Omega_k$ can be defined more efficiently with $\alpha$-shapes (concave hulls). In Figure \ref{fig:MontereyBay_LocatingMissingPoints} (b) an $\alpha$-shape is used to wrap $\Omega_k$ (the green points) and defines the boundary $\partial \Omega_d = \partial (\Omega_k \cup \Omega_m)$, which can then be used to distinguish locations outside the coverage domain ($\Omega_o$, blue points) from missing data deemed inside the coverage domain ($\Omega_m$, red points). The $\alpha$-shapes for a given domain are not unique and depend on a parameter that determines how strict the concave hull encloses the domain. The overall computational complexity of finding an $\alpha$-shape for a set of points in two dimensions is $\mathcal{O}(n \ln n)$ where $n = | \Omega_k|$ is the number of points of $\Omega_k$. This computational efficiency enables one to recompute the $\alpha$-shape at each data time frame in case of diurnal data where the data gaps frequently change in time. Additional details on obtaining $\alpha$-shapes is explained in Appendix \ref{sec:AlphaShape}. After finding an $\alpha$-shape, an extra check may be needed to exclude part of the $\alpha$-shape intersecting land. An example of such intersection can be seen on the south side of Monterey Bay in Figure \ref{fig:MontereyBay_LocatingMissingPoints}.

\begin{figure}[!h]
    \centering
    \footnotesize{
    \includegraphics[width=0.95\textwidth]{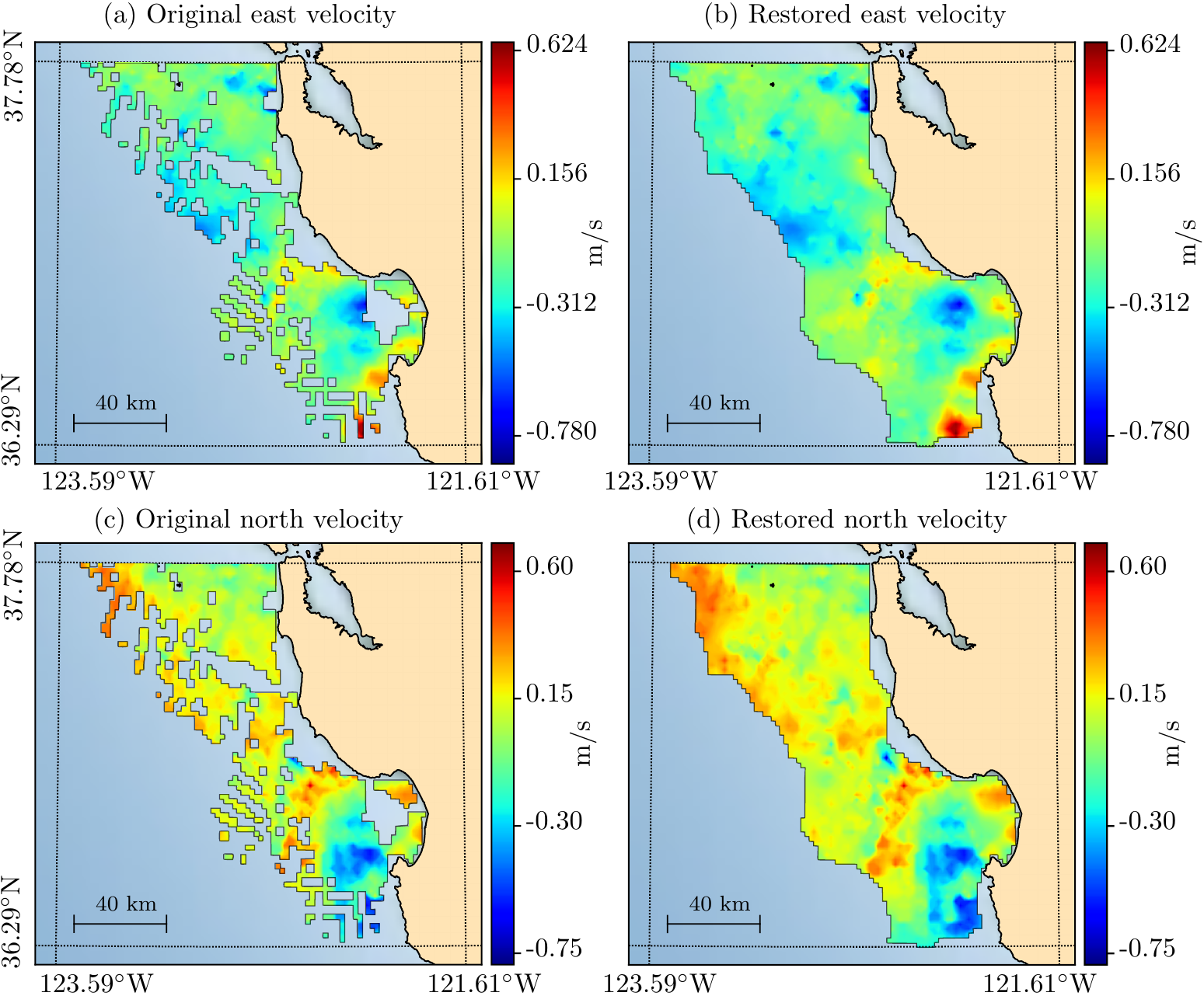}
    }
    \caption{Comparison of original (left) and restored (right) CORDC HF radar velocity data for January 24\textsuperscript{th} 2017, 15:00 UTC.}
    \label{fig:MontereyBay_Velocities}
\end{figure}

The east and north velocity fields for the unprocessed data at January 24\textsuperscript{th} 2017, 15:00 UTC are shown on the left column of Figure \ref{fig:MontereyBay_Velocities}. These fields from the restored data are shown on the right column of Figure \ref{fig:MontereyBay_Velocities}. An inflation size $d = 10$ for the boundary band was used, however, the results were robust to the variation of $d$. The results appear qualitatively accurate.


\subsubsection{Extending restoration to the coast} \label{sec:extend-coast}

Often land-based HF radars provide limited coverage in close proximity to coastlines because, although these regions are adjacent to the radars, the radial signals are too closely aligned to effectively triangulate the velocity components. It is at locations where the radar beams become more orthogonal that better recovery of the velocity can be achieved. The effect of radar placement on measurement error can be characterized by GDOP~\citep{CHAPMAN-1997,CHAPMAN-1997-2,GRABER-1997-2}. In particular, east and north velocity error fields are obtained by a scalar multiplier of the east and north GDOP fields, and often areas with GDOP higher than a threshold (e.g., 1.5) are removed, which can contribute to the spatial gaps in HF radar data.

\begin{figure}[!htpb]
    \centering
    \footnotesize{
    \includegraphics[width=\textwidth]{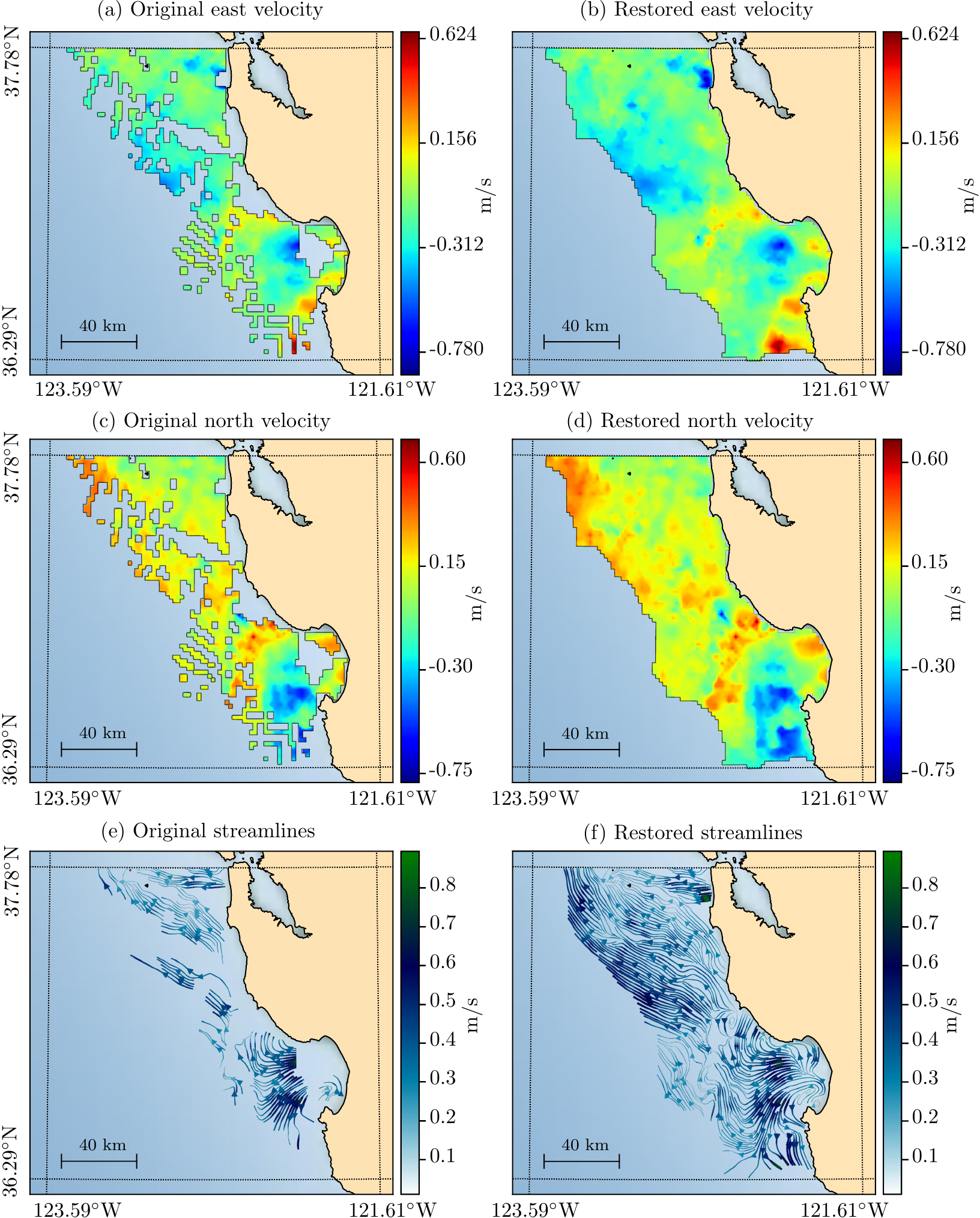}
    }
    \caption{The velocity data of Figure \ref{fig:MontereyBay_Velocities} are restored in a domain that extends to the coastline. Right column is the restored data and left column is the original data that are shown for comparison. First and second rows are east and north velocity fields in order. Third row is the streamlines colored with velocity magnitude.}
    \label{fig:MontereyBay_Land_Velocities}
\end{figure}

In Figure \ref{fig:MontereyBay_LocatingMissingPoints} (b) we have shown the location of radar sites along northern California coastline with their code names. As an example, it can be seen that areas between the two sites in Montara, CA (with the code name MONT) and Pescadero, CA (with the code name PESC) are missing due to the filtering of high GDOP. In Figure \ref{fig:MontereyBay_LocatingMissingPoints} (b) we have extended the missing domain to the coastlines by including the pink points to $\Omega_m$. This can be achieved by finding an $\alpha$-shape around $\Omega_k \cup \Omega_l$ where $\Omega_l$ is the land domain shown in brown points. $\Omega_l$ can be identified by locating the position of each of the points with respect to appropriate coastline data such as the Global Self-consistent, Hierarchical, High-resolution Geography Database \citep{WESSEL-1996}. Since the velocity magnitudes near the coastlines are expected to be smaller than the open boundaries, we have applied no-slip boundary condition on $\partial \Omega_l$ for $\psi$, which leads to zero east and north velocities on $\partial \Omega_l$. The results of restoring the velocity to include near-coastline gaps are shown in the right column of Figure \ref{fig:MontereyBay_Land_Velocities} and can be compared with the original data on the left column of the figure.

%



\section{Discussion} \label{sec:conclusion}

A method for restoring the spatial gaps in oceanographic field data has been presented, which is based on employing transport equations to advect and diffuse field data features into missing regions. This method was applied to two HF radar datasets of varying fidelity. Based on it robust coverage, the MVCO was used primarily for quantitative validation whereby measurements were systematically reduced and replaced by restoration. Excellent quantitative and qualitative agreement was achieved between the restoration and original measurements, with restoration error similar in magnitude to the measurement error intrinsic to the data. The restoration method was also applied to CORDC data from northern California with limited and patchy coverage. The restoration of the CORDC data demonstrated excellent coherency of field patterns even in locations with a large proportion of missing data. 


In prior applications of HF radar restoration, interpolation has been performed by fitting to globally defined bases functions. In such methods, the known field data at measured locations are often altered to fit the model. In the approach presented herein the measured data remains unaltered, and only missing measurements are replaced. Also, in a global approach, several bases functions may be needed to resolve small features, whereas feature size does not generally change the computational cost of the restoration method presented herein. Another quality of the method herein is the ability to control the extent of local information transfer through the prescription of a boundary band size. This achieves a more gradual exchange of field information compared to local interpolation, and more focal data usage compared to a global approach. Lastly, some restoration methods are not objective and depend on the choice of the coordinate system. The PDE based restoration method presented here is independent of translation or rotation of coordinate system making it objective. It can be solved using curvilinear of other coordinates for datasets that span broader scales and are more effectively represented on spherical or manifold surfaces.

{For applications such as obtaining the velocity vector field from radial HR radar measurements, the restoration process can be applied either directly to the radial measurement data, or the reconstructed velocities. Either approach may be considered favorable depending on the application, data, or reconstruction process. For instance, restoration of radial data is not affected by GDOP, although the effect of GDOP errors will eventually emerge once the velocity vector field is reconstructed from the radial data. Such approach would require cylindrical reformulation of PDEs applied to each radial scalar field measurement. On the other hand, if the restoration is performed on the re-gridded velocities after the velocities are reconstructed from the radial data, the Cartesian formulation can be conveniently used over the whole domain, but the process should be applied for each velocity component seperately.

For most of the US coastlines, ocean surface current data are available in real-time from web-based resources, such as the National HF Radar Network. We have developed a web-based gateway (\url{http://transport.me.berkeley.edu/restore}) as a community resource for the restoration of HF radar data. The online gateway can process datasets remotely by the user providing a URL of the data in NetCDF format, which is widely used among the oceanographic community \citep{REW-1990}. The sample data used in this work are provided and the results presented here can be produced and visualized using this online gateway.

In forthcoming work we study the uncertainty quantification and the propagation of errors of the PDE based restoration method that has been presented here. Further possibilities to extend the current work may include spatio-temporal restoration of data where the time correlation is significant, processing and validation of other oceanographic quantities beyond the ocean surface velocity field data, and extension to three-dimensional fields for oceanographic or atmospheric applications.

\begin{appendices}
\section{$\alpha$-shapes} \label{sec:AlphaShape}

$\alpha$-shapes (or concave hulls) are defined in \citep{EDELSBRUNNER-1983}. We briefly describe the algorithm used in our application. Let $\Omega_d$ in Figure \ref{fig:AlphaShape} denote a concave set that is approximated by the finite set $V$ of discrete vertices $v_i = (x_i,y_i) \in V$. The goal is to find the boundary $\partial \Omega_d$. The problem of finding a concave hull around a finite set of vertices does not have a unique solution. For instance in Figure \ref{fig:AlphaShape} by including the edge that connects the two vertices $v_4$ and $v_5$ to $\partial \Omega_d$ a different hull shapes is obtained.

A Delaunay triangulation is performed on $V$ to generate a set of triangles $T$. Each triangle $t \in T$ is a triplet of vertices $t = (v_i,v_j,v_k)$, as shown with dashed lines in Figure \ref{fig:AlphaShape}. The boundary of the union of all triangles $T$ is the convex hull around domain $\Omega_d$. Let $s_i$ denote the circle circumscribing triangle $t_i$, and let $r_i$ denote its radius. Let $T_{\rho} \subseteq T$ be the subset of triangles with $r_i < \rho$, where $\rho \geq 0$ is a threshold to exclude large triangles. The boundary of the union of triangles $T_{\rho}$ defines a concave hull corresponding to the parameter $\rho$.

Clearly $T_0 = \emptyset$ and $T_{\infty} = T$, with the later defining the maximal concave hull, which is also a convex hull. With an adjustment of $\rho$, a desired concave hull can be achieved. If the set of vertices are structured on a rectangular grid with grid spacings $\Delta x$ and $\Delta y$, the smallest meaningful $\rho$ for which $T_{\rho} \neq \emptyset$ is half of the hypotenuse of the smallest triangle with $\rho_{min} = \sqrt{(\Delta x)^2 + (\Delta y)^2}/2$. Practically an $\mathcal{O}(1)$ multiple of $\rho_{min}$ produces a desirable concave hull for our applications.

The main computationally intensive part of finding concave hull is obtaining the Delaunay triangulation. The divide and conquer algorithm is one efficient implementation of Delaunay triangulation with the computational complexity of $\mathcal{O}(n \ln n)$ for $n$ vertices on two dimensional plane.

\begin{figure}[!htpb]
    \centering
    \footnotesize{
    \includegraphics[width=.4\textwidth]{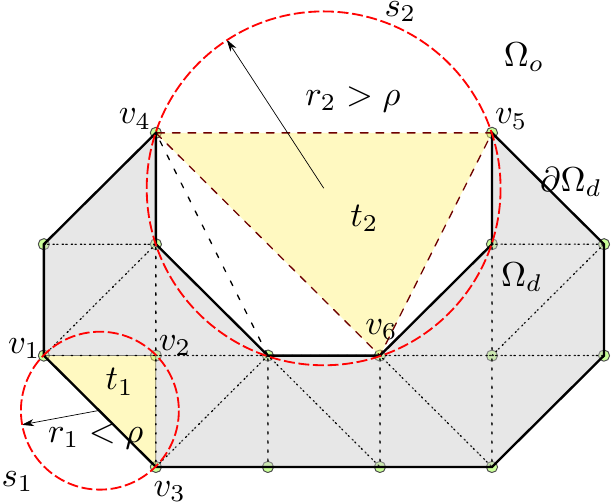}
    }
    \caption{The process of finding an $\alpha$-shape around a set of points $v_i$ is shown. Delaunay triangulation of vertices $v_i$ create triangles $t_i$ that are shown with dashed lines. The radius $r_2$ of circumcircle $s_2$ of triangle $t_2$ is greater than the threshold $\rho$, hence $t_2$ is excluded from the triangles that form the $\alpha$-shape. In contrast, radius $r_1$ of circumcircle $s_1$ of triangle $t_1$ is less than the threshold, so $t_1$ is considered to be one of the triangles that form the $\alpha$-shape. The solid curve $\partial \Omega_d$ is the target $\alpha$-shape.}
    \label{fig:AlphaShape}
\end{figure}

\end{appendices}



\noindent\textbf{Acknowledgement.}
We thank Anthony Kirincich at Woods Hole Oceanographic Institution for providing the Martha's Vineyard HF radar data.
The northern California HF radar data are provided by Coastal Observing Research and Development Center (\url{http://cordc.ucsd.edu/projects/mapping/}). This work was supported by the National Science Foundation, award number 1520825, ``Hazards SEES: Advanced Lagrangian Methods for Prediction, Mitigation and Response to Environmental Flow Hazards''. We thank anonymous reviewers for their constructive suggestions.





\end{document}